\documentstyle[aps,twocolumn]{revtex}
\input psfig

\begin{document}


\draft

\title{Improved Magnetic Information Storage using Return--Point Memory}
\author{Olga Perkovi\'{c} and James P. Sethna}
\address{Laboratory of Atomic and Solid--State Physics,\\
Cornell University,\\
Ithaca,  NY 14853--2501.}
\maketitle

\begin{abstract}

The traditional magnetic storage mechanisms (both analog and digital)
apply an external field signal $H(t)$ to a hysteretic magnetic material,
and read the remanent magnetization $M(t)$, which is (roughly) 
proportional to $H(t)$.  We propose a new analog method
of recovering the signal from the magnetic material, making use of the
shape of the hysteresis loop $M(H)$.  The field $H$, ``stored'' in a
region with $N$ domains or particles, can be recovered with fluctuations
of order $1/N$ using the new method --- much superior to the $1/\sqrt{N}$
fluctuations in traditional analog storage.

\end{abstract}

\pacs{PACS numbers: 85.70.Li, 75.60.Ej, 85.70.Ay, 75.60.-d}

\subsection{Introduction}

How can one best store a song on a magnetic tape?  The traditional
analog method converts the sound signal into a magnetic field $H(t)$,
and then uses it to magnetize the tape, being pulled at a velocity $v$.
The remanent magnetization $M$ (the magnetization left on the tape after
the field has dropped to zero) is roughly linear in $H$:
\begin{equation}
M(x) = M(v t) = C_1\,H(t) + C_2\,H^2(t) + \eta(x).
\label{Nonlinear}
\end{equation}
Here $C_2$ represents the nonlinearity of the remanent magnetization at
high fields (when recording, one turns down the gain until the needle
during the loudest sections stops moving into the red), and $\eta$
represents noise (one turns up the gain as far as possible so quiet
portions don't hiss). (Actually, there are nonlinearities between the
remanent magnetization and the signal $H(t)$ for low fields as well (not
shown in equation\ (\ref{Nonlinear})). When real magnetic tapes are
recorded, the signal $H(t)$ is convolved with a high-frequency, large
amplitude signal \cite{McGraw,Davies,Della_Torre,Mee} in order to remove
these distortions (ac--biasing). This however does not affect the new
method for analog storage that we propose.)

Two other excellent methods have been developed to cope with the noise
and nonlinearity in the remanent magnetization. (There are other sources
of noise in a magnetic recording and reading process, {\it e.g.}, 
interference, electronic noise, and head noise
\cite{Della_Torre,Hoagland,Mallinson}. In this paper, we will address
only the noise relevant to the magnetic material: the magnetic noise.)
Dolby noise reduction does a nonlinear transformation of H(t) to boost
quiet sections and dampen loud sections: the inverse transformation is
applied at playback. Digital recordings are even more effective.  The
signal can be (linearly and accurately) encoded as a stream of bits, and
these bits can be recorded and reproduced without noise or distortion.

How does the analog method compare to digital recording, in terms of the
amount of information that one can store on a given piece of magnetic
tape? One important\cite{McGraw,Della_Torre,Mallinson} source of noise
is the lumpiness of the irreversable magnetization changes in the
material.  Magnetic tapes are often made up of single-domain particles;
if the field is strong enough, some particles will rotate their
magnetizations to the crystallographic axis closest to the direction of 
the field\ \cite{Jiles}.  Other materials with large domains will
magnetize through the depinning of sections of their domain walls,
which (roughly) jump from one pinning center to another.  For our
purposes, these details aren't crucial --- we mostly care about the
number $N$ of these pinning centers or particles.  We will refer
to these lumps, imprecisely, as {\it domains}.


Having more, smaller domains leads to a higher information density.
Averaging over $N$ domains will reduce the fluctuation in the average
magnetization by a factor of $1/\sqrt{N}$ (presuming the interactions
between domains are not important). Thus, the number of different values
of $H$ that can be distinguished in the remanent magnetization scales
like $\sqrt{N}$.  If we subdivide the slice into $Q$ portions, and
magnetize each portion separately, we can store $\sqrt{N/Q}^Q$ different
signals.  Optimizing, we find $Q=N/e$, and we store $e^{N/2\,e}$
distinguishable signals. This is precisely what makes the binary
``digital'' recording so effective: $2^Q$ strings of $Q$ bits are
stored, and our formula suggests that four domains can store one bit
($Q=N/4$).  Of course, substantial error correction would be needed in
order to keep the accuracy high at this scale!

There are times when one is stuck with an analog signal.  Important
recordings have been made with these outdated methods (Beatles' masters)
and potentially one might want to reconstruct signals imposed by natural
processes (reconstructing the stress history of a plastically deformed
material).  We show here that one can do substantially better than the
traditional analog retrieval, by using the portion of the magnetization curve
$M(H)$ near the applied field $H_{signal}$ ({\it e.g.}, by the 
tape head during recording),
rather than just the remanent magnetization at zero field $M(0)$.
We discuss the advantages within the context of two models: the
Preisach\cite{PM,Mayergoyz} model of non-interacting hysteretic domains
(which despite its simplistic assumptions is a standard
tool\cite{PreisachTool} in the engineering community), and the
zero--temperature random--field Ising model (RFIM) \cite{Sethna,RFIM} (a
more realistic model incorporating nearest-neighbor couplings between
domains with different threshold fields).  In the end, we improve our
resolution by $\sqrt{N}$ (from the $1/\sqrt{N}$ resolution of the
remanent magnetization to $1/N$), which for a typical slice of magnetic
tape with $N=10^6$ domains per wavelength (see conclusion) produces a
large improvement in fidelity.  Our method also suffers much less from
nonlinearity: although the fidelity decreases at large magnetizations,
the signature tracks the applied field directly. There is a drawback to
our new method though: by measuring the response curve, the
original signal is necessarily erased.

We should mention another method for dealing with the random noise in
magnetic films that has been developed recently \cite{Economist} by Des
Mapps of Plymouth University in England. Instead of the usual two heads
used in recording (one for demagnetizing the magnetic material and the
other to record the signal), a third head is added, which reads the
signal right after it has been recorded, and sends it to a computer for
analysis. Since the computer ``knows'' what the initial signal was, it
can adjust for the inherent noise in the magnetic material, record the
now ``modulated'' signal, and leave a signature of what it has done.
This provides the information to the ``reading'' head of how to
compensate, during the reading process. Similar techniques have been
independently developed by Indeck and Muller from Washington University
in St. Louis \cite{Economist}.

\subsection{Hysteresis Loops, Subloops, Kinks, and Return--Point Memory}

\begin{figure}
\centerline{
\psfig{figure=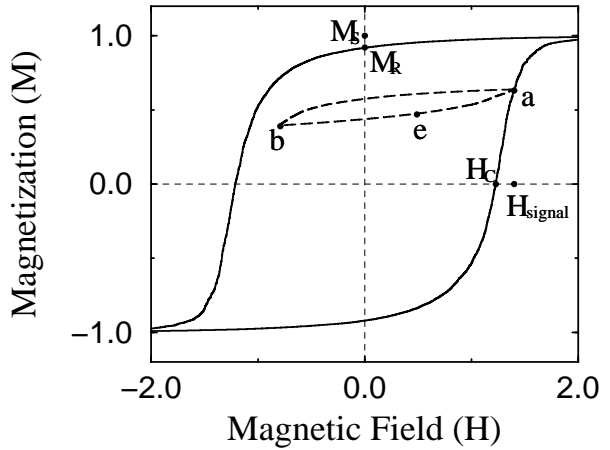,width=3.5truein}}
\caption{ Magnetization $M$ of a ferromagnet as a function of the external field
        $H$.
        The hysteresis curve appears due to a lag between the magnetization
        and the field. $M_S$ denotes the saturation magnetization, $M_R$ the
        remanent magnetization, and $H_C$ the coercivity (see text). The
        subloop $a \rightarrow b \rightarrow e \rightarrow a$ shows the
        return--point memory. The system comes back to the same state $a$,
        as the external field is switched off and then back on.}
\label{hysteresis_fig}
\end{figure}

We review briefly the various kinds of $M(H)$ hysteresis curves relevant
to our discussion.  (There are many different curves, of course, since
the response depends on the magnetic history of the material.) A
ferromagnetic material has the property that its magnetization $M$ lags
behind the external magnetic field $H$, as the field is changed (fig.\
\ref{hysteresis_fig}). This is called hysteresis (which means to lag or
fall behind).  The largest magnetization the material can have (by
aligning all the magnetic domains in the direction of the external field
$H$) is called the saturation magnetization $M_S$. When the field is
switched off, the remaining magnetization is the remanent magnetization
$M_R$, while the field necessary to bring the magnetization to zero is
called the coercivity $H_C$. Variations in the values of these
properties in different ferromagnets make magnetic materials useful for
different applications. For example, magnetic recording materials have
high remanence and coercivity to prevent unwanted demagnetization\
\cite{Jiles}. Therefore, magnetic materials used in recording will in
general have ``square'' hysteresis loops.  By sweeping from very small
to very large magnetic fields, one explores this saturated, ``outer''
hysteresis loop $M_{outer}^{\pm}$ (also called the major hysteresis
loop): any other field history will typically be discussed in terms of
subloops (or minor hysteresis loops) (see fig. \ref{hysteresis_fig}).

\begin{figure}
\centerline{
\psfig{figure=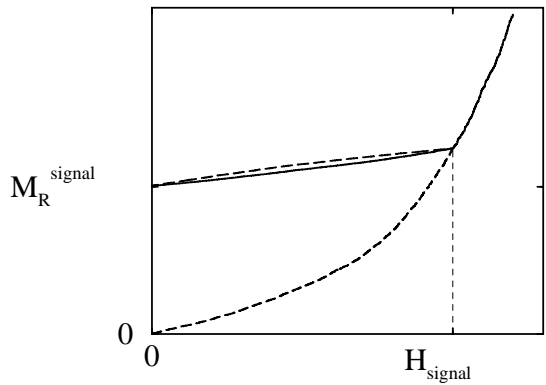,width=3.5truein}}
\caption{The external field is increased to $H_{signal}$ and then decreased
        to zero (dashed curve).
        In the traditional analog storage, the information is
        stored in the remanent magnetization $M_R^{signal}$. In the new
        method, which uses the return--point memory property, the information
        is stored in the field $H_{signal}$ itself, read by increasing
        the external field from zero and finding a ``kink'' in the
        $M$--$H$ curve (solid curve).}
\label{traditional_and_new}
\end{figure}

Before recording, the magnetic tape is demagnetized.  This involves
imposing a slowly decaying, oscillatory field $H(t)$ which leaves the
material in a well-defined, reproducible state with no remanent
magnetization (up to the noise).  Analog recording takes us from this
particular demagnetized state to a magnetized state under an external
field $H_{signal}$: in real recording, this is done by adding an
oscillating field to the signal (see introduction), but initially
we will consider a monotonically increasing field, leading to an
increasing magnetization $M_{imprint}(H)$.
Releasing the external field, the magnetization $M_{relax}$ again drops,
but to a non-zero remanent $M_R^{signal}$ (figure\
\ref{traditional_and_new}). Measuring $M_R^{signal}$ gives information about
$H_{signal}$.

We will also be interested in $M_{measure}(H)$ curve, formed by starting
from the magnetized state $M_R^{signal}$ and raising the field again. As
one sees from figure\ \ref{traditional_and_new}, as the external field
is decreased from the original signal $H_{signal}$ and then increased
again, the magnetization forms a subloop.  As the external field passes
$H_{signal}$, there will be very generally a kink in the
$M_{measure}(H)$ curve (figure\ \ref{traditional_and_new}). This follows
in a direct way, for example, for models exhibiting {\sl return point
memory} \cite{footnote}
(also known as {\sl wiping out}\ \cite{Mayergoyz}). 

For these
systems, the subloop closes exactly: the system returns at $H_{signal}$
to precisely the same state it was in at the peak of its recording field
(wiping out all information about the excursion to lower fields). The
curve $M_{measure}(H)$ above $H_{signal}$ thus necessarily extends
smoothly the original curve $M_{imprint}(H)$, while below $H_{signal}$
it disagrees with $M_{imprint}$: hence it must have a non-analyticity at
$H_{signal}$. Magnetic recording materials often wipe out rather well:
decreasing the field from $H_{signal}$ (and hence repeating the loop)
will reproduce the same subloop (including even the noise) to a good
extent\cite{Urbach}.  On the other hand, magnetic materials --- especially
those like spin-glasses with important antiferromagnetic couplings ---
can exhibit ``reptation'', where repeated cycles lead to a slow
drift in magnetization\cite{Reptation}.
Many other systems (e.g. martensites prepared with
parallel twin boundaries \cite{Martensites}, helium capillary condensing
in a porous material \cite{Hallock}, and superconductors in external
magnetic fields \cite{superconductors}) can exhibit the return--point
memory to various extents; see the work by Amengual {\it et al.}
\cite{Martensites} for an experiment reproducing incredible fine
structure within repeated loops.  Reference \cite{Sethna} discusses
three conditions \cite{footnote} (partial ordering, no passing, and adiabaticity)
which suffice to produce a perfect return-point memory.  The models
studied here possess these properties: antiferromagnetic couplings
violate ``no passing''.

For the purposes of this paper, the kink in $M_{measured}(H)$ is easily
explained.  As the field is raised a second time, the domains which
flipped upon the first rise during imprinting and which did not flip
back upon relaxing are not active.  When $H$ crosses $H_{signal}$, new,
``virgin'' domains are explored: more domains will flip per unit field,
leading to a discontinuity in the slope. It is precisely this slope
discontinuity in the magnetization curve which we propose to use in
information storage. Since the slope discontinuity is at $H_{signal}$,
we are saved from the nonlinearity and distortion which occurs with only
measuring the remanent magnetization.  We will see in the next section
that we also suppress the noise greatly.

\subsection{The Preisach and Random Field Ising Models}

\begin{figure}
\centerline{
\psfig{figure=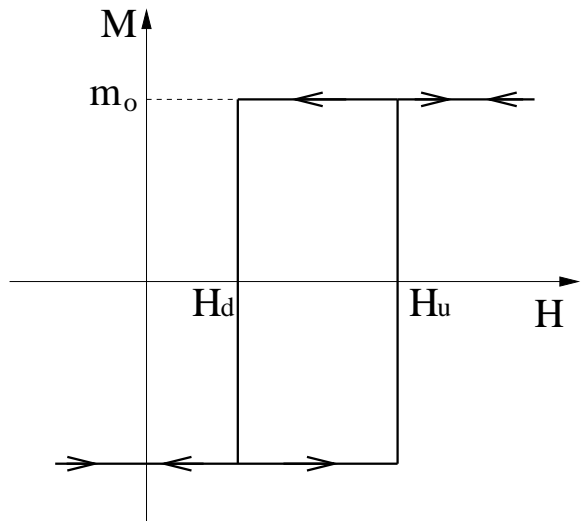,width=3truein}}
\caption{Hysteresis loop for one domain in the Preisach model. The field at
        which the domain flips up is $H_u$ and the field at which it flips
        down is $H_d$. A superposition of many such domain hysteresis
        gives a hysteresis curve as in figure\ \protect\ref{hysteresis_fig}.}
\label{domain_hyst}
\end{figure}

The comparison between the ``traditional'' magnetic analog storage
method, in which the magnetization carries the information, with the new
method where the whole hysteresis loop is retrieved, and the information
is stored directly in the magnetic field, is done using the Preisach and
random field Ising models. The Preisach model\ \cite{PM,Mayergoyz}
consists of a system of independent domains, each with an upper field
$H_u$ and a lower field $H_d$, at which the domain changes sign (or
direction). Each domain therefore has its own square hysteresis loop
(see figure\ \ref{domain_hyst}), and the superposition of all these
loops gives a hysteresis loop as in figure\ \ref{hysteresis_fig}. In
general, the upper and lower fields are not equal and opposite, but the
magnetization per domain $m_0$ is assumed constant. The Preisach model
has the necessary properties for its hysteresis loop to exhibit
return--point memory and is quite useful because of its simplicity, but
since there are no interactions between the domains, the modeling of
real systems is more limited. The random field Ising model
\cite{Sethna,RFIM}, on the other hand, includes nearest neighbor
interactions.

\begin{figure}
\centerline{
\psfig{figure=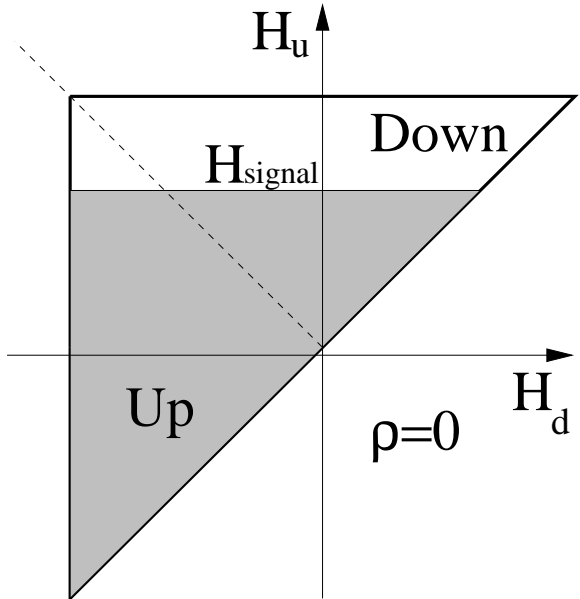,width=3truein}}
\caption{The triangular region is the region in $(H_d, H_u)$ space where
        the weighting function $\rho$ is
        defined (outside of it, $\rho=0$).
        The shape of this region
        is in practice much more general, and depends on the
        properties of the magnetic material\ \protect\cite{Woodward},
        but the region is usually symmetric around the $H_u=-H_d$ line.
        As the field is raised from a large negative value of the field
        to $H_{signal}$ (see fig.\
        \protect\ref{hysteresis_fig}), the domains in the shaded region flip up.
        The magnetization is obtained from equation\
        (\protect\ref{magnetization_eqn}).}
\label{rho_shaded}
\end{figure}

\begin{figure}
\centerline{
\psfig{figure=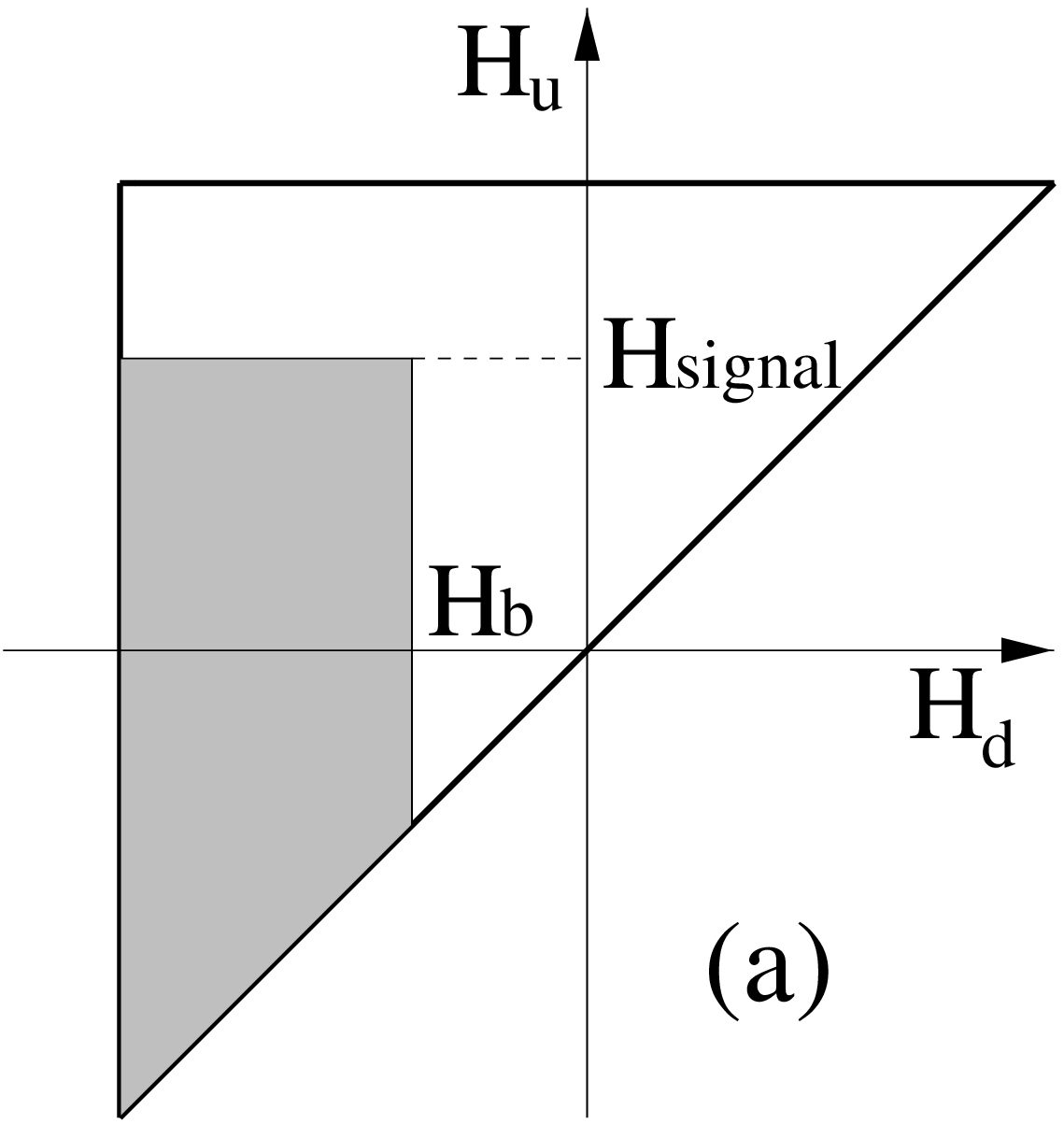,width=3truein}}
\centerline{
\psfig{figure=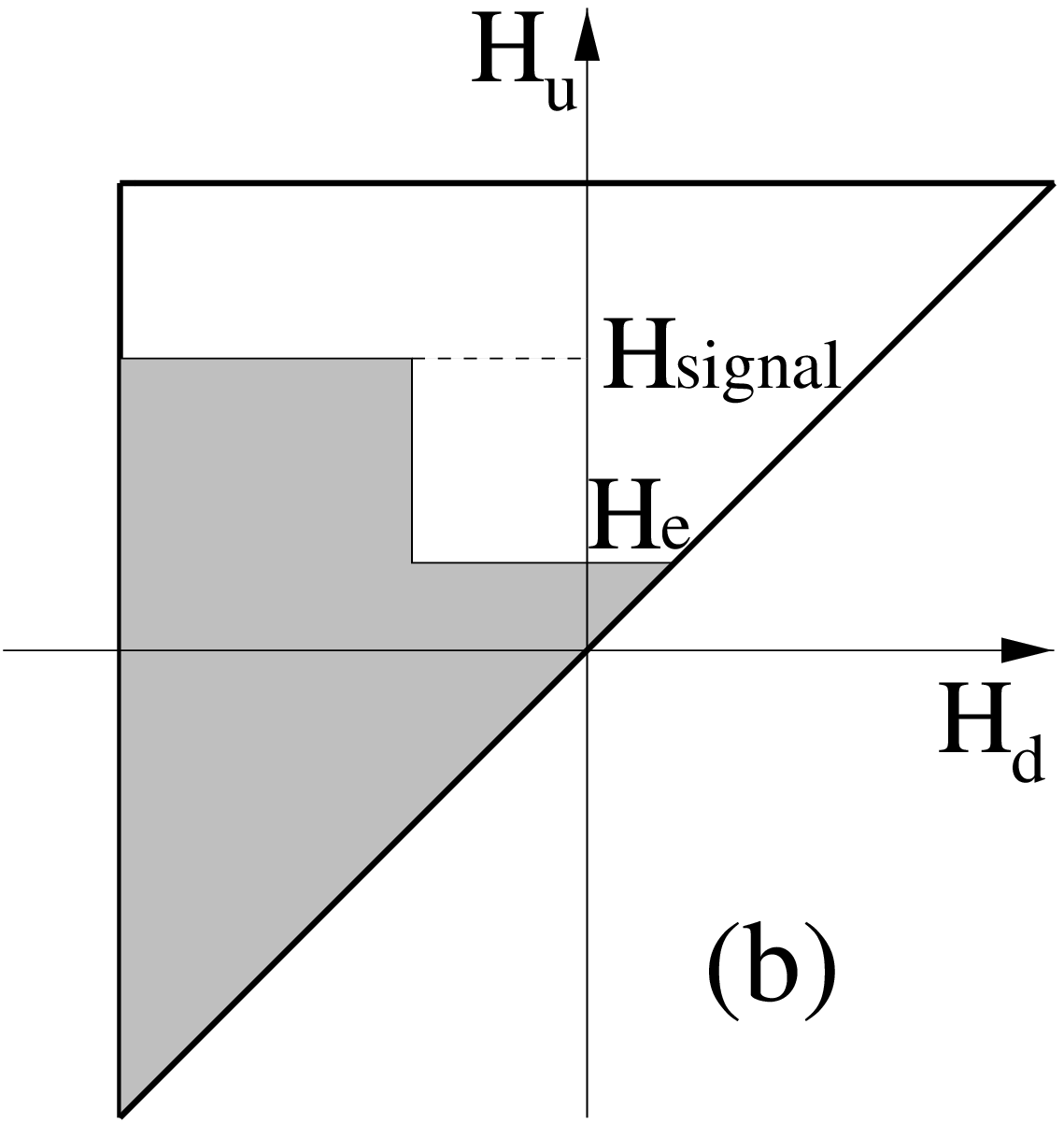,width=3truein}}
\caption{(a) The $(H_d,H_u)$ plane after the field has been increased to
        $H_{signal}$ (from a large negative value), and then decreased to
        a value $H_b$ (see fig. \protect\ref{hysteresis_fig}).
        The shaded area corresponds to spins (domains) that have
        flipped (and not flipped back).
        (b) The field has now been increased to a value $H_e$. More spins
        have flipped, and we find a ``step'' in the Preisach plane. Several
        subloops would give a ``staircase'' (see for example figure
        \protect\ref{H_demagnetize}b).}
\label{staircase_fig}
\end{figure}

We first review the elements of the Preisach model that we will need. To
calculate the magnetization in the Preisach model, a weight function
$\rho (H_d, H_u)$ ($H_u \ge H_d$) needs to be defined. $\rho$ has the
units of magnetization per unit field squared. For systems with
time--reversal symmetry, $\rho$ is symmetric around the line $H_u=H_d$
in the $(H_d, H_u)$ plane\ \cite{PM,Mayergoyz} (see figure\
\ref{rho_shaded}). If all the domains are pointing down, and we increase
the external field from a large negative value to $H_{signal}$ in
figure\ \ref{hysteresis_fig}, the domains whose field $H_u$ fall in the
shaded area of figure\ \ref{rho_shaded} will flip up. The magnetization
of the system is then given by:
\begin{equation}
M = \int\!\!\int_U \rho (H_d, H_u)\ dH_u dH_d -
		 \int\!\!\int_D \rho (H_d, H_u)\ dH_u dH_d
\label{magnetization_eqn}
\end{equation}
where the first integration is over the area where the domains are
pointing up (U), and the second is over the area where the domains are
pointing down (D). If the system follows the path depicted in figure\
\ref{hysteresis_fig} (from a large negative field to $H_{signal}$ to
$H_b$ to $H_e$), we obtain a ``step'' of flipped spins in the
$(H_d,H_u)$ plane (figures\ \ref{staircase_fig}(a,b)). (Several subloops
would give a ``staircase''.) The magnetization is again obtained using
equation\ (\ref{magnetization_eqn}).

\begin{figure}
\centerline{
\psfig{figure=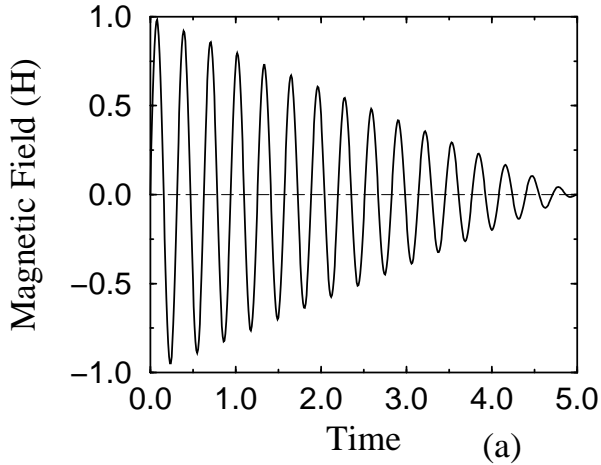,width=3.5truein}}
\centerline{
\psfig{figure=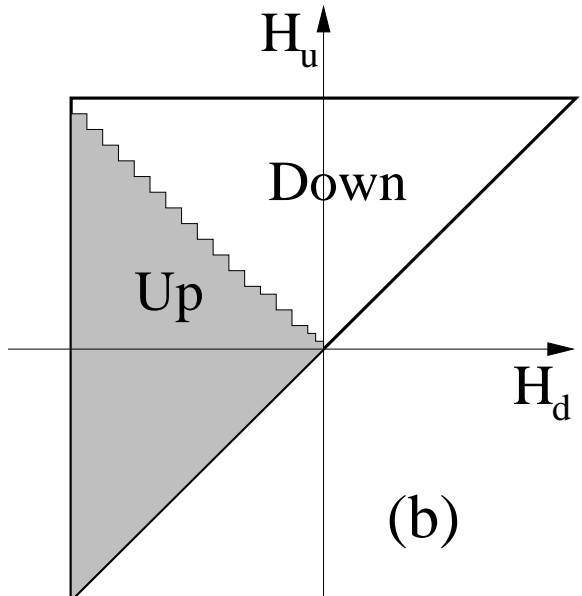,width=3truein}}
\caption{(a) The ac magnetic field used to demagnetize the tape before storing
        the analog signal.
        (b) The demagnetization process in the $(H_d,H_u)$ plane. The
        ``staircase'' occurs after taking the system through smaller and
        smaller loops.}
\label{H_demagnetize}
\end{figure}

The zero--temperature random field Ising model has the following
Hamiltonian\ \cite{Sethna,RFIM}:
\begin{equation}
{\cal H} = - \sum_{<i,j>} J_{ij} s_i s_j - \sum_{i} (H + h_i) s_i
\label{model_equ1}
\end{equation}
where $J_{ij}$ is the nearest neighbor interaction between spins
(domains) $s_i$ and $s_j$ (we set all $J_{ij} = J = 1$), $H$ is the
uniform external magnetic field, and $h_i$ is a random field at site
$s_i$ given by a Gaussian probability distribution. The dynamics is such
that a spin $s_i$ will flip when its ``effective'' field $h^{ef\!f}_i =
J \sum_{j} s_j + H + h_i$ changes sign. A spin that flips can trigger
other spins to flip due to the interaction between nearest neighbors:
avalanches of spins are possible. This model gives rise to a hysteresis
loop and has the return--point memory property\ \cite{Sethna}.

We use the Preisach model with a constant magnetization $m_0$ per
domain, to calculate the relative fluctuations in the signal for the
``traditional'' analog magnetic storage and the new method. The weight
function $\rho$ can be written as $N\,m_0\,\tilde \rho$ with $\tilde
\rho$ being the probability distribution for the domains in the
$(H_d,H_u)$ plane. We also simulate magnetic systems of different sizes
using the random field Ising model. These are then compared to the
analytical results.

\subsection{Traditional Analog Magnetic Storage}

In the traditional analog recording, the tape (system) is first
demagnetized\ \cite{Woodward} by applying a strong ac field which is
gradually reduced to zero (figure\ \ref{H_demagnetize}a). In figure\
\ref{H_demagnetize}b, the demagnetization process is show in the $(H_d,
H_u)$ plane for the Preisach model. In the limit of a very fine
``staircase'' (the ac field needs to drop off very slowly), the
magnetization of the system become zero.

\begin{figure}
\centerline{
\psfig{figure=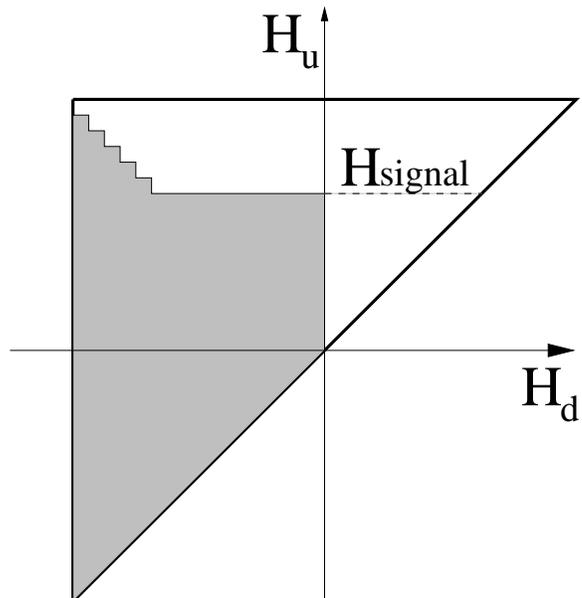,width=3truein}}
\caption{Recording the signal as the remanent magnetization $M_R^{signal}$,
        starting from a demagnetized system (in the Preisach Plane). The
        shaded area represents the system at $M=M_R^{signal}$ and $H=0$.
        The average value of the remanent magnetization $M_R^{signal}$ can
        be calculated using equation (\protect\ref{magnetization_eqn}).}
\label{analog_fig}
\end{figure}

The tape is now ready to be recorded. The external field is raised from
zero until a value $H_{signal}$ (below the saturation value of the
system), and then switched off. At zero field, the tape stays magnetized
with the remanent magnetization $M_R^{signal}$ (figure\
\ref{traditional_and_new}). The magnetization $M_R^{signal}$ can be calculated
from figure\ \ref{analog_fig} using equation\
(\ref{magnetization_eqn}). However, we can calculate the magnetization
differently if we notice that the probability for a domain to be
pointing up is given by:
\begin{equation}
p=\int\!\!\int_U \tilde \rho(H_d,H_u)\, dH_u\,dH_d
\label{prob_p}
\end{equation}
where the integration is over the shaded region in figure\
\ref{analog_fig}. For $N$ independent domains, the probability
$P(n;N,p)$ for observing $n$ up domains out of a total of $N$ is given
by the binomial distribution\ \cite{Binomial}:
\begin{equation}
P(n;N,p)= { N! \over {n!\,(N-n)!}}\,p^{n}\,(1-p)^{N-n}
\label{binom_dist}
\end{equation}
Then, the average number of domains that are up is $Np$, with a root
mean square (rms) of $\sqrt{Np\,(1-p)}$.

Since we are interested in the measurement of the magnetization, the
average magnetization will be $(2Np-N)\,m_0$. If we take the rms to be
our measure of the fluctuation around the average, then the fluctuation
relative to the saturation magnetization is:
\begin{equation}
{|\Delta M_R^{signal}| \over |M_S|} = {m_0\,2\sqrt{N\,
p(1-p)} \over N\,m_0} =
{2\,\sqrt{p(1-p)} \over \sqrt{N}}
\label{rel_fluctuations}
\end{equation}
where $p$ has a value between zero and one. The size $1/\sqrt N$ of this
relative fluctuation limits the amount of information that can be stored
using the traditional analog magnetic storage, in a system with $N$
domains.

\subsection{New Method for Analog Magnetic Storage}

With the new method for analog storage, the information is stored in the
value of the external field $H_{signal}$ at which the kink in the
hysteresis occurs. Similarly to the traditional analog storage, we have
a limitation on how well the information can be retrieved given by a
value proportional to $\Delta H_{signal}/ H_C$, where $\Delta
H_{signal}$ gives the fluctuation around the value $H_{signal}$, and
$H_C$ is the coercivity.

\begin{figure}
\centerline{
\psfig{figure=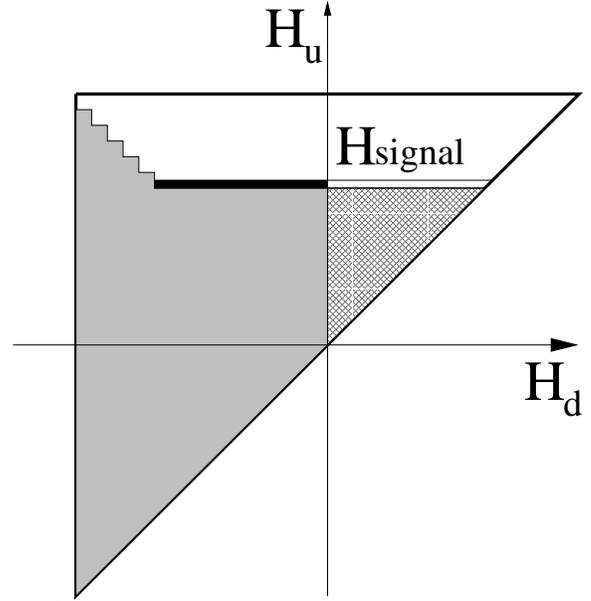,width=3truein}}
\caption{Storage of information in the field $H_{signal}$
        using the return--point memory effect, as seen in the Preisach plane.
        The shaded area represents the system at $M=M_R^{signal}$ and $H=0$.
        As the field
        is increased, spins in the cross--hatched region flip up.
        At $H_{signal}$
        there is a discontinuity in the number of spins flipping per unit field.
        This appears as a kink in the $M$--$H$ curve. The integral of
        the weight function $\rho$ over the black horizontal
        strip is ($f^{(2)}-f^{(1)}$), and over the white strip is $f^{(1)}$.}
\label{rpm_fig}
\end{figure}

\begin{figure}
\centerline{
\psfig{figure=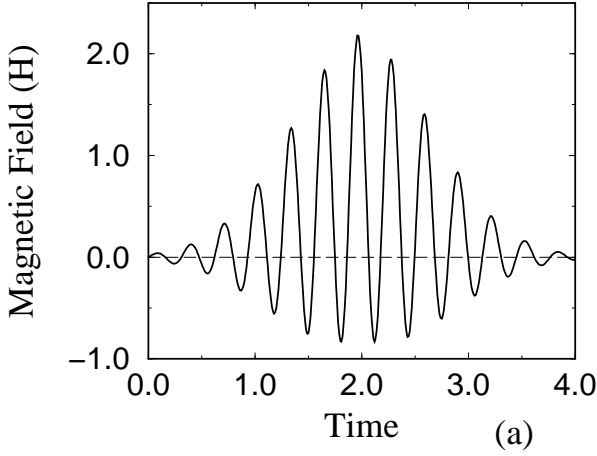,width=3.5truein}}
\centerline{
\psfig{figure=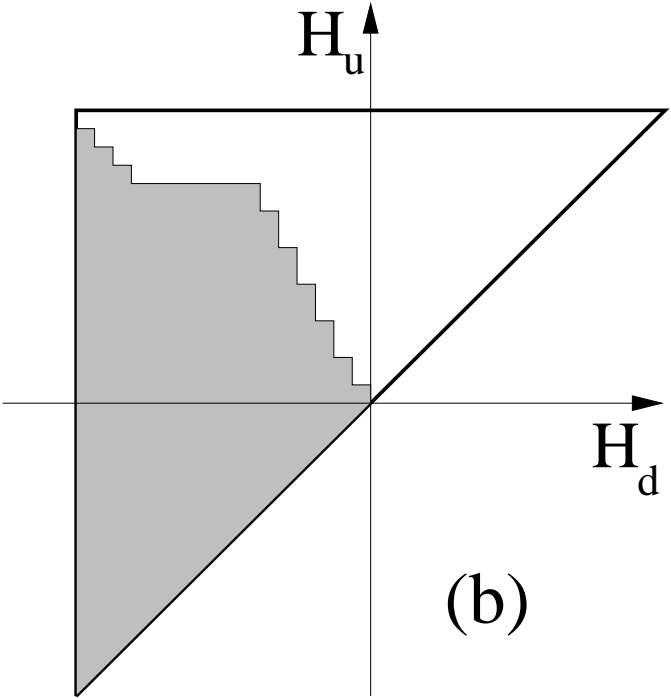,width=3.5truein}}
\caption{(a) Magnetic field as ``seen'' by a magnetic tape moving past
        a recording head with an ac--biased
        field superimposed on a constant field $H_{signal}$.
        (b) The Preisach plane after the ac--biased signal in (a)
        has been ``stored''.
        If the field is now increased (from $H_U=0$), there is
        a discontinuity in the number of domain flips per unit field as
        we pass the ``large'' step. That value of the field corresponds
        to approximately $H_{signal}$ shifted by the amplitude of the
        ac--bias.}
\label{rpm_acbias}
\end{figure}

As before, we start with the tape (system) demagnetized, and increase
the field up to a value $H_{signal}$, smaller than the saturation field.
The external field is then switched off (fig.\
\ref{traditional_and_new}). The system will be magnetized with a
remanent magnetization $M_R^{signal}$. When the information needs to be
recovered, instead of ``reading off'' the value of the remanent
magnetization $M_R^{signal}$, the external field is increased until a
kink in the $M$--$H$ curve is found. If the system exhibits the
return--point memory property, the field at which the kink occurs is
$H_{signal}$. In the $(H_d,H_u)$ plane (figure\ \ref{rpm_fig}), the kink
is seen as a discontinuous increase in the number of domains flipped as
the field is increased past $H_{signal}$. (If the original signal is
ac--biased, as in audio signals, there is a discontinuity in the
Preisach plane \cite{Mee} at a field shifted up from $H_{signal}$ by
about the amplitude of the ac--bias (see figures\ \ref{rpm_acbias}(a,b).
The retrieval of this value is otherwise analogous to the presentation
that follows.)

To observe the kink, we can take the derivative of the hysteresis curve
with respect to the field $H$ and observe the discontinuity. In general
the data will need to be smoothed over some range $\Delta H$, which will
help in finding the discontinuity in the slope, but will introduce
fluctuations around $H_{signal}$ of the order of $\Delta H$. The
discontinuity in the slope can be observed if the difference between the
number of domains that flip in a range $\Delta H$, below and above
$H_{signal}$, is larger than the standard deviation in the number of
domains flipped in $\Delta H$ above $H_{signal}$:
\begin{equation}
\Delta N_{\uparrow}^{(2)} - \Delta N_{\uparrow}^{(1)} >
			\sqrt {Np^{(2)}(1-p^{(2)})}
\label{difference_eqn}
\end{equation}
where the superscripts $(1)$ and $(2)$ indicate measurements below and
above $H_{signal}$ respectively, and $p^{(2)}$ is the probability that a
domain flips from down to up in a range $\Delta H$, just above
$H_{signal}$. (In general we should require that the difference be
larger than the fluctuations just above {\it and} below $H_{signal}$,
but since the fluctuations below $H_{signal}$ are smaller, we can use
equation\ (\ref{difference_eqn}).) Note that $Np^{(2)}= \Delta
N_{\uparrow}^{(2)}$. If the interval $\Delta H$ is small enough, that
the slope $f=dM/dH$ of the $M$--$H$ curve can be considered close to
constant, we have:
\begin{equation}
{(f^{(2)}-f^{(1)}) \over m_0}\, \Delta H > \sqrt {{f^{(2)} \over m_0}\,
\Delta H\, (1-p^{(2)})}
\label{difference_eqn2}
\end{equation}
Thus the uncertainty in the measured $H_{signal}$ is:
\begin{equation}
\Delta H_{signal} \sim \Delta H >
{m_0\, f^{(2)}\, (1-p^{(2)}) \over (f^{(2)}-f^{(1)})^2}
\label{difference_eqn3}
\end{equation}

From figure\ \ref{rpm_fig}, in the $(H_d,H_u)$ plane, $f^{(2)}$ is
$N\,m_0 \int_{-H_{signal}}^{H_{signal}} \tilde \rho(H_d,H_{signal})\
dH_d$ and the difference $(f^{(2)}-f^{(1)})$ is $N\,m_0
\int_{-H_{signal}}^0 \tilde \rho(H_d,H_{signal})\ dH_d$. Then, the
fluctuation in the field $H_{signal}$ relative to the coercivity $H_c$
is:
\begin{equation}
{|\Delta H_{signal}| \over |H_c|} \sim
{1 \over N}\
{(1-p^{(2)})\, \int_{-H_{signal}}^{H_{signal}}
\tilde \rho(H_d,H_{signal})\ dH_d \over
|H_c|\, \Bigl(\int_{-H_{signal}}^0 \tilde \rho(H_d,H_{signal})\ dH_d\Bigr)^2}
\label{H_fluct_eqn3}
\end{equation}
The ratio multiplying $1/N$ in equation\ (\ref{H_fluct_eqn3}) is of
order one as long as the signal is not too small. For small
$H_{signal}$, this ratio diverges since in the Preisach model near
$(M=0,H=0)$, the $M$--$H$ curve is quadratic, and the difference
$(f^{(2)}-f^{(1)})$ between the two slopes is negligible. This
divergence can be avoided if the signal is stored after the system has
been saturated (instead of starting with a demagnetized system).
Therefore, away from $(M=0,H=0)$, the ``number'' of fields that can be
used to store information scales as $N$.

\subsection{Random field Ising Model Simulation Results}

In the previous two sections, we have obtained the scaling with the
system size $N$ of the relative fluctuations in the magnetization (for
the traditional storage method) and the field (for the new storage
method). The analysis was done for independent domains (spins). We now
simulate the storing and reading process for both methods, using the
random field Ising model, which includes nearest neighbor interactions.

For the traditional storage method, we increase the field up to a value
$H_{signal}$ (from a large negative value) and then turn it off.
We then measure the average magnetization $M(H=0)$ for up to one
hundred initial random field configurations, and measure the standard
deviation $\Delta M_R^{signal}$. We define the relative fluctuation as
$\Delta M_R^{signal} / M_R$, where $M_R$ is the remanent
magnetization, and is equal to about $0.92$ (within $4\%$) for a
disorder of $R=3$ (recall that $R$ is the standard deviation of the
Gaussian distribution of random fields).

\begin{figure}
\centerline{
\psfig{figure=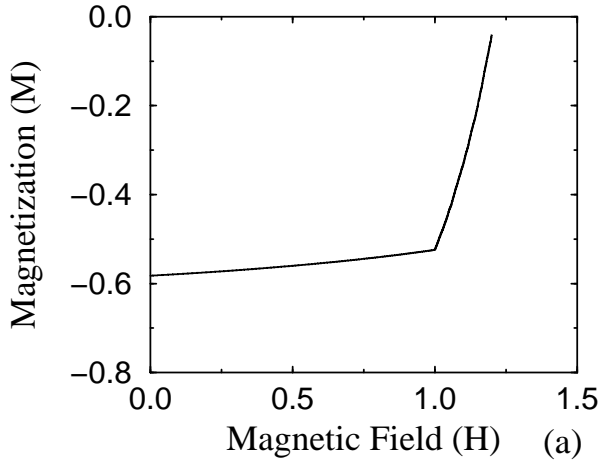,width=3.5truein}}
\centerline{
\psfig{figure=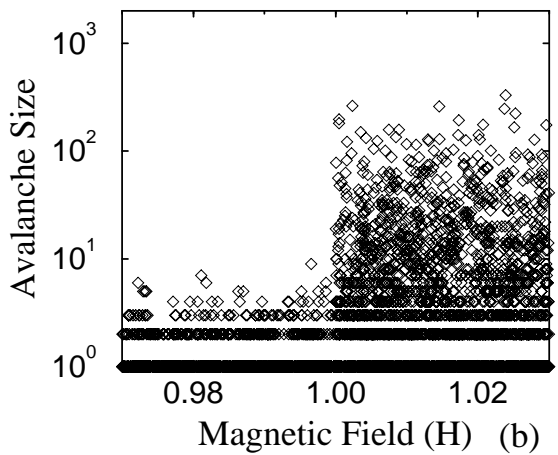,width=3.5truein}}
\caption{(a) Simulation of the reading process for the new method of
        analog storage, for a $100^3$ system size and disorder $R=3$.
        The kink in the $M$--$H$ curve corresponds to the field $H_{signal}$.
        (b) Number of spins flipping at the field $H$ (avalanche size)
        for the data in (a) near $H_{signal}$ (which is one). The
        field $H_{signal}$ is found when a threshold avalanche size (here $13$)
        is reached.}
\label{simul_reading}
\end{figure}

With the new method, we store the information the same way, but instead
of reading off $M(H=0)$, we increase the external field until a kink in
the magnetization curve is found. The field at which the kink occurs
should be the field $H_{signal}$. Figure\ \ref{simul_reading}a shows the
reading process. We define the relative fluctuation as the difference
between the field read off, $H_{read}$, and the ``real'' field,
$H_{signal}$, divided by the coercivity $H_C$ (which is about $1.21$ for
$R=3$).

To find $H_{read}$, we note that the smaller slope of $dM/dH$ inside the
subloop reflects the smaller size of the jumps in the magnetization, or
avalanches. Therefore the field $H_{read}$ is the field at which some
``threshold'' magnetization jump (or avalanche size) is reached
(fig.~\ref{simul_reading}b).

\begin{figure}
\centerline{
\psfig{figure=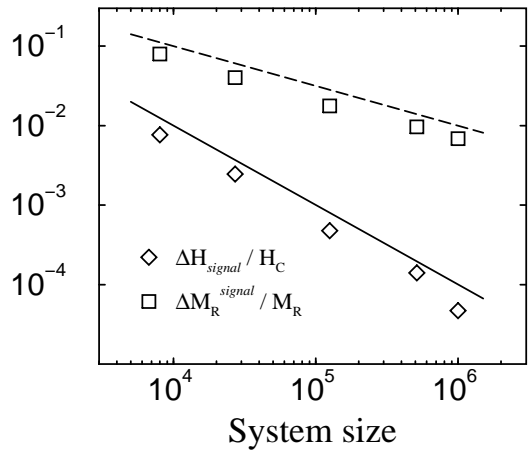,width=3.5truein}}
\caption{ Relative fluctuations for the field (diamonds) and the magnetization
       (square), at several system sizes. The solid line shows a $1/N$
       behavior, and the dashed line a  $ N^{-1/2}$ behavior, where
       $N$ is the system size. }
\label{delta_H_M_fig}
\end{figure}

Figure \ref{delta_H_M_fig} shows the results of our simulation. The
diamonds correspond to the relative fluctuations in the field as defined
above, with a threshold of $13$ spins in an avalanche. Note that the
behavior follows the $1/N$ scaling (solid line), while the relative
fluctuations in the magnetization (squares) follow the $1/{\sqrt N}$
scaling (dashed line). The simulation was done for $20^3$, $30^3$,
$50^3$, $80^3$, and $100^3$ spins. The figure suggests a crossover at a
system size of $100$ spins, below which the relative fluctuations for
the magnetization will become smaller than for the field.

\subsection{Summary and Conclusion}

We have shown, using the Preisach model of non-interacting domains, that
the new method of analog storage which uses the property of
return--point memory, gives fluctuations in the signal that are smaller
than the ones found for the traditional method. The difference is
approximately given by a factor of $\sqrt N$, where $N$ is the number of
domains in the system being magnetized. The same behavior is found in
our simulation of a magnetic system with nearest neighbor interactions
and randomness. A question remains: how large is $N$ for typical
magnetic tapes?

For analog storage, we can estimate the number of domains per cycles.
The typical ferromagnetic grain sizes found in particulate media
magnetic tapes are $0.5\,\mu m$ in length and $0.1\,\mu m$ in diameter\
\cite{Jiles,Middelhoek,White}, and the area covered by one grain is
about $5 \times 10^{-10}\,cm^2$. The grains used in magnetic recording
are usually too small to contain a domain wall and can therefore be
considered as single domain particles\ \cite{Jiles}. The packing on the
tape is usually less than $40\%$. If we assume the percentage to be
$35\%$, then the surface grain (domain) density is $7 \times 10^8$
grains per $cm^2$. The magnetic tapes are typically $1.28\,cm$ wide
($1/2''$), and therefore in $1\,cm $ (in length) of tape, there are
about $9 \times 10^8$ grains. For typical consumer tapes, the speed at
which the tape is moved is close to $5\,cm/s$\ \cite{McGraw,Middelhoek}.
Therefore, the number of grains (domains) per cycle is $7.5 \times 10^7$
for a $60\,H\!z$ signal, and $2 \times 10^5$ for a $20\,kH\!z$ signal.
Professional tapes have speeds of up to $76\,cm/s$ and the number of
grains $N$ per cycle is $1 \times 10^9$ and $3 \times 10^6$
respectively, for the two frequencies. Therefore, since $N$ is large, a
$\sqrt N$ drop in the signal fluctuation is quite significant. (For
digital storage, the recording densities are as large as $20$ million
bits per $cm^2$ \cite{Bertram,digital_noise}, which for a
polycrystalline thin film medium gives $\sim 5000$ grains per bit of
information, for a $10^{-3}\,\mu m^2$ grain size.) As for the erasure of
the stored information as it is retrieved, the added fidelity and
linearity should compensate for having to rewrite the tape after it is
read.


We acknowledge the support of NSF Grant \#DMR-9419506. We would like to
thank Sivan Kartha who wrote the code for the simulation, and Bruce W.
Roberts, David M. Goodstein, James A. Krumhansl, and Karin A. Dahmen for
helpful conversations. This work was conducted on IBM 560 workstations
(donated by IBM). We would like to thank the Material Science Computer
facility, and IBM for their support.

%


\begin{references}
%
\bibitem{McGraw} McGraw--Hill Encyclopedia of Science and Technology:
	{\it Magnetic recording}, 7th ed. (McGraw--Hill, 1992).
%
\bibitem{Davies} S. J. Begun, {\it Magnetic Recording} (Rinehart \& Company,
	NY, 1955); G. L. Davies,
	{\it Magnetic Tape Instrumentation} (McGraw--Hill, New York, 1961);
	B. B. Bycer, {\it Digital Magnetic Tape Recording: Principles and
	Computer Applications} (Hayden Book Company, NY, 1965).
%
\bibitem{Della_Torre} E. Della Torre, {\it Magnetic Recording}, Encyclopedia of
        Physical Science and Technology, Vol. 9 (Academic Press, 1987);
%
\bibitem{Mee} C. D. Mee, {\it The Physics of Magnetic Recording}
	(North--Holland Publishing, Amsterdam, 1964).
%
\bibitem{Hoagland} A. S. Hoagland and J. E. Monson, {\it Digital Magnetic
       Recording}, 2nd edition (John Wiley, New York, 1991).
%
\bibitem{Mallinson} J. C. Mallinson, {\it The Foundations of Magnetic
	Recording} (Academic Press, San Diego, 1987).
%
\bibitem{Jiles} D. Jiles, {\it Introduction to Magnetism and Magnetic
         Materials} (Chapman and Hall, New York, 1991).
%
\bibitem{PM} F. Preisach, Z. Phys. {\it 94}, 277 (1935).
%
\bibitem{Mayergoyz} I. D. Mayergoyz, {\it Mathematical Models of Hysteresis}
        (Springer--Verlag, Berlin, 1991).
%
\bibitem{PreisachTool} I. D. Mayergoyz, J. Appl. Phys. {\bf 57} (1), 3803
	(1985); J. Ort\'in, J. Appl. Phys. {\bf 71} (3), 1454 (1992), and
	J. Phys. IV, Colloq. {\bf 1} C4-65 (1991).
%
\bibitem{Sethna} J. P. Sethna, K. A. Dahmen, S. Kartha, J. A. Krumhansl,
        B. W. Roberts, and J. D. Shore, Phys. Rev. Lett. {\bf 70}, 3347 (1993);
%
\bibitem{RFIM} K.~A.~Dahmen, S.~Kartha,
	J.~A.~Krumhansl, B. W. Roberts,
        J. P. Sethna, and J. D. Shore, J. Appl. Phys. {\bf 75} (10),
        5946 (1994); K. A. Dahmen and J. P. Sethna, Phys. Rev. Lett. {\bf 71},
        3222 (1993); K. A. Dahmen and J. P. Sethna, accepted for publication
        in PRB;
        O.~Perkovi\'c, K.~A.~Dahmen, and J. P. Sethna, Phys. Rev.
        Lett. {\bf 75}, 4528 (1995); O. Perkovi\'c, K. A. Dahmen, and
	J. P. Sethna, submitted to PRB.
%
\bibitem{Economist} {\it The Economist}, March 5th 1994, page 94.
%
\bibitem{footnote} Theoretically, the return--point memory effect can be
explained by using the ``No Passing'' rule introduce by Middleton in the
study of sliding density waves\ \protect\cite{Middleton}, and by assuming
adiabaticity\ \cite{Sethna}. Let's define a state ${\bf s}=s_1,...,s_N
\ge {\bf r}=r_1,...,r_N$ if $s_i \ge r_i$ for every site $i$ in the
system. Then the ``No Passing'' rule says that if a system ${\bf s}(t)$
evolves under the field $H_s(t)$, and another system ${\bf r}(t)$ under
$H_r(t)$, and we have the conditions ${\bf s}(0)>{\bf r}(0)$ and $H_s(t)
> H_r(t)$ for all times, then ${\bf s}(t)>{\bf r}(t)$ for all times as
well. Adiabacity means that the field changes slowly enough that if the
system starts in some state ${\bf s_a}$, {\it any monotonic} path from a
field $H_a$ to a field $H_b$ will bring the system to the same state
${\bf s_b}$. Using these two properties, Sethna {\it et al.} have proven
the existence of return--point memory (see\ \cite{Sethna} for the
proof).
%
\bibitem{Middleton} A. A. Middleton, Phys. Rev. Lett. {\bf 68}, 670 (1992);
        A. A. Middleton and D. S. Fisher, Phys. Rev. B {\bf 47}, 3530 (1993).
%
\bibitem{Urbach} J. S. Urbach, R. C. Madison, and J. T. Markert, Phys.
        Rev. Lett. {\bf 75}, 4694 (1995).
%
\bibitem{Reptation} L. N\'eel, J. de Phys. Rad. {\bf 20}, 215 (1959);
L.~P. L\'evy, J.~de Physique I {\bf 3}, 533 (1993).
%
\bibitem{Martensites} A. Amengual, LL. Ma\~{n}osa, F. Marco, C. Picornell,
        C. Segui, and V. Torra, Thermochimica Acta {\bf 116}, 195 (1987).
        J. Ort\'in, Journal de Physique IV, Colloque C4-65 (1991),
        and J. Appl. Phys. {\bf 71} (3), 1454 (1992).
%
\bibitem{Hallock} M. P. Lilly, P. T. Finley, and R. B. Hallock, Phys. Rev.
        Lett. {\bf 71}, 4186 (1993); M. P. Lilly and R. B. Hallock,
        Physica B {\bf 194--196}, 691 (1994).
%
\bibitem{superconductors} G. Friedman, L. Liu, and J. S. Kouvel, J. Appl.
        Phys {\bf 75} (10), 5683 (1994).
%
\bibitem{Woodward} J. G. Woodward and E. Della Torre, J. Appl. Phys. {\bf 31},
        56, 1960.
%
\bibitem{Binomial} P. R. Bevington and D. K. Robinson, {\it Data Reduction
	and Error Analysis for the Physical Sciences}, 2nd ed.
	(McGraw--Hill, New York, 1992); H. J. Larson, {\it Introduction to
	Probability Theory and Statistical Inference}, 3rd ed. (Wiley,
	New York, 1982).
%
\bibitem{Middelhoek} S. Middelhoek, P. K. George, and P. Dekker, {\it
       Physics of Computer Memory Devices} (Academic Press, New York, 1976).
%
\bibitem{White} R. M. White, Sci. Amer. {\bf 243}, 138 (1980); and
	IEEE Spectrum {\bf 20}, 32 (1983).
%
\bibitem{Bertram} H. N. Bertram and J.--G. Zhu, Solid State Physics {\bf 46},
	271, eds. H. Ehrenreich and D. Turnbull (Academic Press,
	San Diego, 1992).
%
\bibitem{digital_noise} For more information on digital noise see also:
	{\it Noise in Digital Magnetic Recording}, eds. T. C. Arnoldussen
	and L. L. Nunnelley (World Scientific, Singapore, 1992).


\end{references}
\end{document}